\documentstyle[11pt,epsfig]{article}

\begin{document}

\title{Loop induced $W^{\pm}H^{\mp}$ associated production via photon-photon
collisions in the THDM and the MSSM \thanks{The project supported by National
           Natural Science Foundation of China}}
\author{{Zhou Fei$^{b}$, Ma Wen-Gan$^{a,b}$, Jiang Yi$^{b}$, Li Xue-Qian$^{a,c}$ and
Wan Lang-Hui$^{b}$} \\
{\small $^{a}$ CCAST (World Laboratory), P.O.Box 8730, Beijing
100080,P.R.China} \\
{\small $^{b}$ Department of Modern Physics, University of Science and
Technology}\\
{\small of China (USTC), Hefei, Anhui 230027, P.R.China} \\
{\small $^{c}$ Department of Physics, Nankai University, Tianjin 300070,
China}}
\date{}
\maketitle

\begin{abstract}
We study the loop induced production of charged Higgs boson
associated with W-boson via photon-photon collisions at linear
colliders (LC). The investigation is carried out under the
two-Higgs-doublet model(THDM) and the minimal supersymmetric
model (MSSM) using full on-mass-shell renormalization scheme.
The numerical analysis of their production rates is presented with
some typical parameter sets. The results show that the total cross sections
in frame of the MSSM at linear colliders(LC) can reach 0.1 femto barn
quantitatively. The contributions from supersymmetric
sector can lead to an order of magnitude enhancement of the cross
section in some parameter space. We conclude that the associated
$W^{\pm}H^{\mp}$ production rate in photon-photon collision mode
is comparable to that in $e^+e^-$ collision mode.
\end{abstract}

\vspace{3mm}

\vskip 12mm

\vskip 5cm

{PACS number(s): 12.60.Jv, 12.60.Fr, 12.15.Lk, 13.85.-t}

\vfill \eject

\baselineskip=0.26in

\renewcommand{\theequation}{\arabic{section}.\arabic{equation}} %
\renewcommand{\thesection}{\arabic{section}}

\makeatletter      
\@addtoreset{equation}{section} \makeatother

\section{Introduction}
\par
As we know the Higgs sector is one of the essential aspects of the
Standard Model(SM) \cite{s1}\cite{s2}, responsible for spontaneous
symmetry breaking and the generation of masses for the fermions
and gauge bosons. But until now the origin of electroweak symmetry
breaking has not yet been directly tested by experiments. The
extensions of the SM can lead to additional physical Higgs bosons.
Particularly the general two-Higgs-doublet model(THDM) and the
minimal supersymmetric model(MSSM) \cite{s3}\cite{haber}
necessarily involve charged Higgs bosons $H^{\pm}$. Because of its
electrical charge, the charged Higgs boson is noticeably different
from neutral Higgs particles. Therefore, experimental discovery of
Higgs bosons including charged Higgs bosons is one of the
important goals of the present and future colliders.
\par
Here we review briefly the production mechanisms for charged Higgs bosons
at hadron colliders for a comparison between the production
mechanisms at LC and at the TeV energy scale hadron colliders,
At hadron colliders, there are several mechanisms which can
produce charged Higgs bosons\cite{Djou}.
\newline
(1) The charged Higgs boson pair can be produced by $q\bar{q} \rightarrow
H^{+}H^{-}$ \cite{Desh} and gluon-gluon fusions at one-loop $gg \rightarrow
H^{+}H^{-}$ \cite{Will} at the LHC. Since the heavy $H^{\pm}$-bosons decay
dominantly into quark pairs, the pair production process is always
accompanied by serious QCD backgrounds.
\newline
(2) When $m_{H^{\pm}} < m_t-m_b$, the $H^{\pm}$ bosons can abundantly be
produced in decays of top (anti-top) quarks $t\bar{t} \rightarrow bH^{+} (%
\bar{b}H^{-})$ from the parent production channel $pp\rightarrow t\bar{t}$.
The dominant decay channels in this mass range are $H^{+}(H^{-}) \rightarrow
\bar{\tau}\nu_{\tau} (\tau \bar{\nu}_{\tau})$.
\newline
(3) When $m_{H^{\pm}} > m_t-m_b$, the single charged Higgs boson production
can be via $gb(g\bar{b}) \rightarrow t H^{-} (\bar{t} H^{+})$ \cite{Gun}, $%
gg \rightarrow t\bar{b}H^{-} (\bar{t} bH^{+})$\cite{Diaz} and $qb(\bar{q}%
\bar{b}) \rightarrow q^{\prime}bH^{+}(\bar{q}^{\prime}\bar{b}H^{+})$
processes \cite{Oda}. The sequential decay $H^{+} \rightarrow t\bar{b}$ is
known as a preferred channel for $H^{\pm}$ boson search. But signals in
these processes appear together with large QCD background. The associated
heavy $H^{\pm}$ production with $W$ boson is another important channel in
production of the heavy $H^{\pm}$ bosons which has been investigated in Refs.
\cite{glwh}\cite{glwh1}\cite{glwh2}. In this case the $W^{\pm}$-boson's
leptonic decay may be used as a spectacular trigger.
\par
In the future linear colliders (LC) there are several mechanisms
which can produce charged Higgs bosons:
\par
(1) Pair production via $e^+e^- \to H^{+}H^{- }$ and $e^+e^-
\to \gamma \gamma \to H^{+}H^{-}$. The former has already been
studied at the tree-level in Ref.\cite{eech1} and one-loop order
in Refs. \cite{eech2}\cite{eech3}. The latter production process
is through laser back-scattered $\gamma\gamma$ collisions and has
been studied in Ref.\cite{Ma1}. It is found that these two processes
are prominent in discovery of the charged Higgs bosons.
\par
(2) Single production of charged Higgs boson in association with $W$
gauge boson at $e^+e^-$ colliders provides an attractive alternative in
searching for $H^{\pm}$, which is kinematically favored when
$m_{H^{\pm}}$ exceeds $m_W$. In Ref.\cite{eewh1} the associated
production mechanism via direct $e^+e^-$ collisions has been
studied in the THDM. Since there is no tree-level
$H^{\pm}W^{\mp}V$ couplings $(V=\gamma$ and $Z^0)$,
both the $H^{\pm}W^{\mp}$ assiociated production processes via
$\gamma\gamma$ and $e^+ e^-$ collisions have no tree-level
contribution in the THDM and the MSSM and occur at one-loop
level in the lowest order.
\par
In this paper we concentrated ourselves on the investigation of the
process $e^+e^- \to \gamma \gamma \to W^{\pm}H^{\mp}$
including the complete contributions in a non-SUSY THDM and the
MSSM at LC. Since both the $W^{\pm}H^{\mp}$ associated production
processes via $e^+e^-$ and laser back-scattered $\gamma\gamma$
collision modes are all one-loop induced at the lowest order,
and from the technical point of view there is no problem in realizing
the luminosity of laser back-scattered photon beam approching to that
of electron beam, the production rate through $\gamma\gamma$ collisions
could be competitive with the $e^+e^-$ colliding mode at LC. We arrange
this paper as follows. In Sec.2 we present the analytical
calculation. In Sec.3 we give some numerical presentations and
discuss the numerical results of the processes $e^+e^- \to \gamma
\gamma \to W^{\pm}H^{\mp} $. In frame of the MSSM we do numerical
calculation in the minimal supergravity (mSUGRA) scenario
\cite{msugra}. The conclusions are contained in Sec.IV. Finally
some notations used in this paper and the explicit expressions of
the self-energies are collected in the Appendix.

\section{The Calculation of $e^+e^- \to \gamma \gamma \to W^{+}H^{-} $}

\par
In this section we investigate the process
\[
e^+e^- \to \gamma \gamma \to W^{+}H^{-}
\]
in frames of a CP conserving non-SUSY THDM and MSSM. In these models the cross
section of $e^+e^- \to \gamma \gamma \to W^-H^+$ coincides with the process
$e^+e^- \to \gamma \gamma \to W^+H^-$ because of the charge conjugation
invariance. Our calculation in this section is concentrated on the process
$e^+e^- \to \gamma \gamma \to W^+ H^- $, unless otherwise stated. But the
numerical results of the total cross section in Section III involve both
two processes i.e. $e^+e^- \to \gamma\gamma \to W^{\pm} H^{\mp} $. Hence our
numerical results of total cross section contain a factor of 2 in contrast
to only $W^+H^-$ production.
\par
In our calculation, we adopt the 't Hooft-Feynman gauge, and use dimension
regularization scheme for the calculations in frame of the non-SUSY THDM and
dimension reduction scheme in frame of the MSSM. We denote the reaction of
$W^{+}H^{-}$ production via photon-photon collisions as:
\begin{eqnarray*}
\gamma(p_1, \mu) \gamma (p_2, \nu) \longrightarrow W^+ (k_1, \lambda) H^-
(k_2).
\end{eqnarray*}
where $p_1,~p_2$ and $k_1,~k_2$ represent the four momenta of the incoming
photons and outgoing $W^+$ and $H^-$, respectively. The Mandelstam variables
are defined as $\hat{s}=(p_1+p_2)^2$, $\hat{t}=(p_1-k_1)^2$ and
$\hat{u}=(p_1-k_2)^2$.
\par
We depict the generic Feynman diagrams contributing to the subprocess
$\gamma \gamma \to W^-H^+$ in Fig.1 in the frame of the MSSM. As shown in the
figure, this is a one-loop induced subprocess. The diagrams with internal
supersymmetric particles $\tilde{q}$, $\tilde{{\chi}^+}$ and $\tilde{{\chi}^0}$
do not appear in the non-SUSY THDM. In Fig.1 there are self-energy,
triangle, box and quartic one-loop diagrams as well as the counter-term
for the $\gamma-W^{+}-H^-$ vertex appearing in t-channel. The
corresponding u-channel diagrams are not shown in this figure.
The existance of vertex counterterm is due to the fact that this
vertex is absent at the tree level, but is UV-divergent at loop
contributions. The $\gamma-W^{+}-H^-$ vertex counter-term can be
visualized in Fig.1(b.1), (b.2). As this subprocess involves
non-diagonal gauge propagators, the loop diagrams in
Fig.1(a.1-a.19) and the one-loop diagrams in Fig.1(b.3-b.5)
are not UV-finite by themselves. Both the mixing $H^+-W^+$
and $H^+-G^+$ propagators have to be renormalized to
obtain finite physical results. The explicit self-energy one-loop diagrams
for mixings of $H^+-W^+$ and $H^+-G^+$ are plotted in Fig.2.
Since the $H^+-W^+$ self-energy diagrams on the external $W$
boson have no contribution as a consequence of on-mass-shell $W$
gauge boson, we do not plot them on Fig.1. By using the
renormalization condition as described in Refs. \cite{dabelstein}
\cite{denner}, the sum of Higgs tadpole diagrams and tadpole
counter-terms vanishes. Therefore, there is no need to consider
the tadpole diagrams in the self-energy calculations.
\par
The Higgs sector of a non-SUSY THDM or the MSSM consists of two
scalar doublets. In reference \cite{Ros} the complete set of
Feynman rules for the MSSM is given. Here we follow their
notations and conventions. We denote the two Higgs doublets as
\begin{eqnarray*}
H_1 = \left(
\begin{array}{c}
H_1^1 \\
H_1^2
\end{array}
\right) = \left(
\begin{array}{c}
(v_1 + \phi_1^{0} - i \chi_1^{0})/\sqrt{2} \\
- \phi_1^-
\end{array}
\right), \ H_2 = \left(
\begin{array}{c}
H_2^1 \\
H_2^2
\end{array}
\right) = \left(
\begin{array}{c}
\phi_2^+ \\
(v_2 + \phi_2^0 + i \chi_2^0)/\sqrt{2}
\end{array}
\right)
\end{eqnarray*}
\par
Theoretically, the self-interactions among charged Higgs bosons of
two-Higgs-doublet extension models and the gauge bosons are completely
determined by local gauge invariance. The kinetic terms read:
\begin{eqnarray*}
{\cal L}_{kin} &=& \left[(\vec{\partial}_{\mu}+i g_1 \frac{Y_{H_1}}{2}
B_{\mu} + i g_2 T^a W^a_{\mu})H_1\right]^{\dag} (\vec{\partial}_{\mu}+i g_1
\frac{Y_{H_1}}{2} B_{\mu} + i g_2 T^a W^a_{\mu})H_1 +
\end{eqnarray*}
\begin{eqnarray}\label{lagrangian}
\left[(\vec{\partial}_{\mu}+i g_1 \frac{Y_{H_1}}{2} B_{\mu} + i g_2 T^a
W^a_{\mu})H_2\right]^{\dag} (\vec{\partial}_{\mu}+i g_1 \frac{Y_{H_1}}{2}
B_{\mu} + i g_2 T^a W^a_{\mu})H_2 .
\end{eqnarray}
where $T^{a}$ are the $SU(2)$ generators. $B_{\mu}$ and $W^a_{\mu}$ denote
the $U(1)_Y$ and $SU(2)_L$ gauge fields respectively, and $g_1$ and $g_2$
correspond to their relevant coupling constants. $Y_{H_i}~(i=1,2)$ are the
hypercharges of Higgs fields. Our one-loop calculation involves $H^+-W^+$
and $H^+-G^+$ mixing self-energies and their relevant counterterms. We adopt
the on-mass-shell renormalization scheme suggested in Ref.\cite{dabelstein}
for the extended Higgs doublets. We define that the Higgs
fields and vacuum expectation values are renormalized as below:
\begin{eqnarray}
H_i &\rightarrow & Z^{1/2}_{H_i} H_i \nonumber\\
v_i &\rightarrow & Z^{1/2}_{H_i}(v_i+\delta v_i)=\left(1+\frac{\delta %
v_i}{v_i}+\frac{1}{2}\delta Z_{H_i} \right) v_i , \label{renorm}
\end{eqnarray}
hence
$$
\frac{\delta \tan{\beta}}{\tan{\beta}}= \frac{\delta v_2}{v_2}- %
\frac{\delta v_1}{v_1} + \frac{1}{2}(\delta Z_{H_2}-\delta
Z_{H_1}).
$$
\par
We replace of the free parameters and the fields in the kinetic
terms of the Lagrangian Eq.(2.1) by the renormalized parameters
and fields to obtain corresponding counterterms at the one-loop
order in the Lagrangian, then we can read out all the relevant
counterterms for our process. We define the counterterms $\delta Z_{HG}$
and $\delta Z_{HW}$, which behave as if they were the renormalized
constants introduced as:
\begin{eqnarray}
\left(
\begin{array}{c}
H^{\pm} \\
G^{\pm}
\end{array}
\right) \rightarrow \left(
\begin{array}{cc}
Z_{H^{\pm}} & \delta Z_{HG}^{(1)} \\
\delta Z_{HG}^{(2)} & Z_{G^{\pm}}
\end{array}
\right) \left(
\begin{array}{c}
H^{\pm} \\
G^{\pm}
\end{array}
\right),
\end{eqnarray}
and
\begin{eqnarray}
W^{\pm}_{\mu} \rightarrow (Z_2^{W})^{1/2} W_{\mu}^{\pm} \pm i\frac{\delta
Z_{HW}}{m_W} \partial_{\mu} H^{\pm},
\end{eqnarray}
where $Z_2^{W}=1+\delta Z_2^{W}$ is the usual $SU(2)_L$ gauge
triplet renormalization constant. Then we get the relations
\begin{eqnarray}
\delta Z_{HG}^{(1),(2)}&=&\frac{\sin{2\beta}}{2}
\left[\frac{1}{2}(\delta Z_{H_2}-\delta Z_{H_1}) \mp \frac{\delta
\tan\beta}{\tan\beta} \right], \nonumber \\
\delta Z_{HG}&=&\delta Z_{GH}^{(1)}+\delta Z_{HG}^{(2)}=
\frac{\sin{2\beta}}{2}(\delta Z_{H_2}-\delta Z_{H_1}), \nonumber \\
\delta Z_{HW}&=& \delta Z_{HG}^{(2)}.
\end{eqnarray}
And the counterterm for the $\gamma WH$ vertex appearing in
Fig.1(b.1,b.2) has the form as
\begin{eqnarray}
\delta V_{\gamma WH} = i e g_{\mu\nu} m_W \delta Z_{HG}^{(2)},
\end{eqnarray}
which is obtained with the above replacement of the Lagragian terms
${\cal L}=e ~m_W A^{\mu} W^-_{\mu} G^+ +{\rm h.c.}+\dots$
in Eq.(\ref{lagrangian}).
\par
Denoting the bare self-energy as $\Sigma$, the renormalized one is
defined as $\hat{\Sigma}=\Sigma-\delta\Sigma$. Here we define the
renormalized two-point Green functions for the $H^{+}-W^{+}$
and $H^{+}-G^{+}$ mixings as
$$
G_{\mu}^{HW}=\frac{-i}{p^2-m_{H^{\pm}}^2}
p^{\nu}(\Sigma_{HW}-\delta\Sigma_{HW})\frac{-ig_{\mu\nu}}{p^2-m_{W}^2},
$$
$$ G^{HG}=\frac{-i}{p^2-m_{H^{\pm}}^2}
(\Sigma_{HG}-\delta\Sigma_{HG})\frac{i}{p^2-m_{W}^2},
$$
respectively. The counterterm $\delta\Sigma$ can be derived from the
bare self-energy expression by using on-mass-shell scheme. With an
on-mass-shell charged Higgs boson, the real part of the renormalized
mixing self-energy $\hat{\Sigma}_{HW}|_{p^2=m_{H^{\pm}}}$ should be zero, i.e.
\begin{eqnarray}
{\it Re}\hat{\Sigma}_{HW}|_{p^2=m^2_{H^+}}=0 \Longrightarrow
\delta Z_{HW} = \frac{{\it Re}\Sigma_{HW}(m_{H^{\pm}}^2)}{m_W^2}.
\end{eqnarray}
\par
For the renormalized self-energies ${\hat{\Sigma}}_{HW}$ and
${\hat{\Sigma}}_{HG}$, we have the Slavnov-Taylor identity \cite{coa}
\begin{eqnarray}
k^2 {\hat{\Sigma}}_{HW}(p^2)-m_W{\hat{\Sigma}}_{HG}(p^2)=0,
~~~~~ at~ p^2=m^2_{H^+}.
\end{eqnarray}
From the above two equations we can get
\begin{eqnarray}
{\it Re}\hat{\Sigma}_{HG}|_{p^2=m^2_{H^+}}=0.
\end{eqnarray}
Then we obtain the related counterterms of the self-energies as
\begin{eqnarray}
\delta \Sigma_{HW}&=& {\rm Re} \Sigma_{HW}|_{p^2=m_{H^+}^2} = {\rm
Re}(
\Sigma_{HW}^{f}+ \Sigma_{HW}^{b})|_{p^2=m_{H^+}^2}, \nonumber \\
\delta \Sigma_{HG}&=& {\rm Re} \Sigma_{HG}|_{p^2=m_{H^+}^2} = {\rm
Re}( \Sigma_{HG}^{f}+ \Sigma_{HG}^{b})|_{p^2=m_{H^+}^2},
\end{eqnarray}
where $\Sigma_{HW,HG}^{f}$ and $\Sigma_{HW,HG}^{b}$ denote the
unrenormalized self-energy parts from the fermion loop and boson loop diagrams,
respectively. The unrenormalized self-energies $\Sigma_{HW}$ and
$\Sigma_{HG} $ can be calculated from the diagrams in Fig.2. Due to the
on-mass-shell renormalization condition,
the quartic self-energy diagrams in Fig.2(5) does not contribute to the
$H^+-G^+$ self-energy.
\par
Since the incoming photons are unpolarized, it is required to average
over their polarization. The sum of the photon
polarization vector $\epsilon^{\mu} (\lambda_{1},p_{1})$ and $\epsilon^{\nu}
(\lambda_{2},p_{2})$ has to be chosen in such a way that only the
physical(transverse) polarization states remain and the unphysical
one(longitudinal) does not contribute to matrix element\cite{been}:
\begin{eqnarray}
\sum_{\lambda=1,2} \epsilon^{\mu *} (\lambda,p) \epsilon^{\nu} (\lambda,p) =
-g^{\mu\nu}+ 2 \frac{p_{1}^{\mu}p_{2}^{\nu}+p_{1}^{\nu}p_{2}^{\mu}}{\hat{s}}%
,~~(p=p_1, p_2).
\end{eqnarray}
The renormalized amplitude of the subprocess is expressed as
\begin{eqnarray*}
{\cal M}&=& {\cal M}_{(t)}^{{\rm loop}} + {\cal M}_{(t)}^{{\rm ct}} + {\cal M%
}_{(t)}^{{\rm bub}} + {\cal M}_{(u)}^{{\rm loop}} + {\cal M}_{(u)}^{{\rm ct}%
} + {\cal M}_{(u)}^{{\rm bub}} \\
&=& {\epsilon^{\mu}(p_1)\epsilon^{%
\nu}(p_2) \epsilon^{\lambda}(-k_1)} \left( f_{1} g_{\mu \nu} p_{2
\lambda} + f_{2} g_{\mu \nu} p_{1 \lambda}+ f_{3} g_{\mu \lambda}
k_{1 \nu}+ f_{4} g_{\nu \lambda} k_{1 \mu}
+ f_{5} k_1^{\alpha} \epsilon_{\alpha \mu \nu \lambda} \right. \\
&+& f_{6} p_1^{\alpha} \epsilon_{\mu \nu \lambda \alpha} + f_{7}
p_2^{\alpha} \epsilon_{\mu \nu \lambda \alpha} + f_{8} k_{1 \mu} k_{1 \nu}
p_{2 \lambda}+ f_{9} k_{1 \mu} k_{1 \nu} p_{1 \lambda}+ f_{10} k_1^{\alpha}
p_1^{\beta} p_{2 \lambda} \epsilon_{\alpha \mu \nu \beta} \\
&+& f_{11} k_1^{\alpha} p_2^{\beta} p_{2 \lambda} \epsilon_{\alpha \mu \nu
\beta} + f_{12} k_1^{\alpha} p_{1 \lambda} p_2^{\beta} \epsilon_{\alpha \mu
\nu \beta} + f_{13} k_1^{\alpha} k_{1 \nu} p_1^{\beta} \epsilon_{\alpha \mu
\lambda \beta} + f_{14} k_1^{\alpha}p_2^{\beta} k_{1 \nu} \epsilon_{\alpha
\mu \lambda \beta} \\
&+& \left. f_{15} k_1^{\alpha}p_2^{\beta} k_{1 \mu}\epsilon_{\alpha \nu
\lambda \beta} + f_{16} p_1^{\alpha}p_2^{\beta} p_{2 \lambda}\epsilon_{\mu
\nu \alpha \beta} + f_{17} p_1^{\alpha}p_2^{\beta} p_{1
\lambda}\epsilon_{\mu \nu \alpha \beta}+ f_{18} k_{1 \nu}
p_1^{\alpha}p_2^{\beta} \epsilon_{\mu \lambda \alpha \beta} \right. \\
&+&\left. f_{19} k_{1 \mu} p_1^{\alpha} p_2^{\beta}\epsilon_{\nu \lambda
\alpha \beta} + f_{20} k_1^{\alpha}p_1^{\beta}p_2^{\gamma} \epsilon_{\alpha
\mu \beta \gamma} g_{\nu \lambda} + f_{21}
k_1^{\alpha}p_1^{\beta}p_2^{\gamma} \epsilon_{\alpha \nu \beta \gamma}
g_{\mu \lambda} \right. \\
&+& \left.f_{22} k_1^{\alpha}p_1^{\beta}p_2^{\gamma}
\epsilon_{\alpha \lambda \beta \gamma} g_{\mu \nu} \right)
={\epsilon^{\mu}(p_1)\epsilon^{\nu}(p_2) \epsilon^{\lambda}(-k_1)}
\sum_{i=1}^{22} f_i {\cal L}_i,
\end{eqnarray*}
where $f_i(i=1,22)$ are the form factors and ${\cal L}_i$ are
their respective Lorentz tensors.
The u-channel matrix parts ${\cal M}_{(u)}^{{\rm loop}}$, ${\cal M}%
_{(u)}^{{\rm ct}}$ and ${\cal M}_{(u)}^{{\rm bub}}$ are obtained by the
following replacements:
\begin{eqnarray*}
{\cal M}_{(u)}^{{\rm loop}} &=& {\cal M}_{(t)}^{{\rm loop}}(t \rightarrow u,
\mu \leftrightarrow \nu, p_1 \leftrightarrow p_2),\\
{\cal M}_{(u)}^{{\rm ct}} &=& {\cal M}_{(t)}^{{\rm ct}}(t \rightarrow u, \mu
\leftrightarrow \nu, p_1 \leftrightarrow p_2),\\
{\cal M}_{(u)}^{{\rm bub}} &=& {\cal M}_{(t)}^{{\rm bub}}(t \rightarrow u,
\mu \leftrightarrow \nu, p_1 \leftrightarrow p_2),
\end{eqnarray*}
The amplitude parts of ${\cal M}_{(t),(u)}^{{\rm loop}}$, ${\cal M}%
_{(t),(u)}^{{\rm ct}}$ and ${\cal M}_{(t),(u)}^{{\rm bub}}$
represent the contributions from the one-loop diagrams
Fig.1(a.1-a.19), the counter term diagrams Fig.1(b.1-b.2) and the
self-energy bubble diagrams Fig.1(b.3-b.5), respectively. We
decompose the amplitude parts of ${\cal M}_{(t)}^{{\rm ct}}$ and
${\cal M}_{(t)}^{{\rm bub}}$ as:
\begin{eqnarray}
{\cal M}_{(t)}^{{\rm ct}}&=& {\cal M}^{(t)}_{b1}+ {\cal M}^{(t)}_{b2},\nonumber \\
{\cal M}_{(t)}^{{\rm bub}}&=& {\cal M}^{(t)}_{b3}+ {\cal M}^{(t)}_{b4} +
{\cal M}^{(t)}_{b5}.
\end{eqnarray}
where
\begin{eqnarray*}
{\cal M}^{(t)}_{b1}&=& {\epsilon^{\mu}(p_1) \epsilon^{\nu}(p_2)
\epsilon^{\lambda}(-k_1)} \frac{2e \delta V_{\gamma WH}}{{\hat t}-m_{H^+}^2} {\cal L}_3 \\
{\cal M}^{(t)}_{b2}&=& {\epsilon^{\mu}(p_1) \epsilon^{\nu}(p_2)
\epsilon^{\lambda}(-k_1)} \frac{e \delta V_{\gamma WH}}{{\hat t}-m_W^2} (2{\cal L}%
_2 - {\cal L}_3 +2{\cal L}_4) \\
{\cal M}^{(t)}_{b3}&=& {\epsilon^{\mu}(p_1) \epsilon^{\nu}(p_2)
\epsilon^{\lambda}(-k_1)} \frac{2i e^2 m_W }{({\hat t}-m_W^2)({\hat t}-m_{H^+}^2)} \left(%
\hat{\Sigma}_{GW} +m_W \hat{\Sigma}_{HW}\right) {\cal L}_3 \\
{\cal M}^{(t)}_{b4}&=& {\epsilon^{\mu}(p_1) \epsilon^{\nu}(p_2)
\epsilon^{\lambda}(-k_1)} \frac{i e^2 m_W }{({\hat t}-m_W^2)(m_{H^+}^2-m_W^2)} \cdot
\\
&& \left\{ \hat{\Sigma}_{GW}(2{\cal L}_2 + {\cal L}_3+ 2{\cal L}_4) +\hat{%
\Sigma}_{HW}\left[2m_W{\cal L}_3+\frac{{\hat t}}{m_W}(2{\cal L}_2 - {\cal L}_3 +2%
{\cal L}_4)\right] \right\} \\
{\cal M}^{(t)}_{b5}&=& {\epsilon^{\mu}(p_1) \epsilon^{\nu}(p_2)
\epsilon^{\lambda}(-k_1)} \frac{-ie^2\hat{\Sigma}_{HW}}{m_{H^+}^2-m_W^2} (2%
{\cal L}_1 +2{\cal L}_2+ {\cal L}_3 +{\cal L}_4),
\end{eqnarray*}
The explicit expressions of the related unrenormalized self-energies are
listed in Appendix.
\par
The cross section for the subprocess at one-loop order with the unpolarized
photon collisions can be obtained from
\begin{eqnarray}
\hat{\sigma}(\hat{s},\gamma \gamma \to W^+ H^-) = \frac{1}{16 \pi \hat{s}^2}
\int_{\hat{t}^{-}}^{\hat{t}^{+}} d\hat{t}~ \bar{\sum\limits_{}^{}}
|{\cal M}|^2.
\end{eqnarray}
In above equation, we define $\hat{t}^\pm=\left[ (m^{2}_H+m^{2}_W -\hat{s}%
)\pm \sqrt{(m^{2}_H+m^{2}_W-\hat{s})^2 -4m^{2}_Hm^{2}_W}\right]/2$. The bar
over the sum means average over initial spins.
\par
With the integrated photon luminosity in the $e^{+}e^{-}$ collision, the
total cross section of the process $e^{+}e^{-} \rightarrow \gamma\gamma
\rightarrow \to W^+ H^- $ can be written as:
\begin{equation}
\sigma(s) = \int_{E_0/\sqrt{\hat{s}}}^{x_{max}} dz \frac{d{\cal L%
}_{\gamma \gamma}}{dz} \hat{\sigma} (\gamma\gamma \rightarrow W^+ H^-
{\rm at\ } \hat{s}=z^2 s)
\end{equation}
with $E_0=m_{W}+m_{H^+}$, and $\sqrt{s}$($\sqrt{\hat{s}}$) being
the $e^{+}e^{-}$($\gamma \gamma$) center-of-mass energy. $d {\cal
L}_{\gamma\gamma}/dz$ is defined as:
\begin{equation}
\frac{d{\cal L}_{\gamma\gamma}}{dz} = 2z \int_{z^2/x_{max}}^{x_{max}} \frac{%
dx} {x} F_{\gamma/e}(x) F_{\gamma/e}(z^2/x).
\end{equation}
\par
For the initial unpolarized electrons and laser photon beams, the energy
spectrum of the back-scattered is given by\cite{sd}\cite{sh}\cite{si}
\begin{equation}
F_{\gamma/e}=\frac{1}{D(\xi)} \left[1-x+\frac{1}{1-x}-\frac{4x}{\xi (1-x)}+
\frac{4x^2}{\xi^2 (1-x)^2} \right],
\end{equation}
where
\begin{eqnarray}
D(\xi)&=&(1-\frac{4}{\xi}-\frac{8}{\xi^2}) \ln (1+\xi)+\frac{1}{2}+\frac{8}{\xi%
} -\frac{1}{2(1+\xi)^2} \nonumber\\
\xi&=&4 E_0 \omega_0 / m_e^2,
\end{eqnarray}
$m_e$ and $E_0$ are the mass and energy of the electron respectively, and $%
\omega_0$ is the laser-photon energy, $x$ represents the fraction of the
energy of the incident electron carried by the back-scattered photon. In out
evaluation, we choose $\omega_0$ such that it maximizes the backscattered
photon energy without spoiling the luminosity through $e^{+}e^{-}$ pair
creation. Then we have $\xi=2(1+\sqrt{2})\simeq 4.8, x_{max}\simeq 0.83$ and
$D(\xi)\simeq 1.8$.

\section{Numerical results and discussions}

\par
In this section, we present some numerical results of the total cross
section for the processes $e^+e^- \to \gamma\gamma \to W^{\pm} H^{\mp}$. The
SM input parameters are chosen as: $m_t=174.3~GeV$, $m_{Z}=91.1882~GeV$, $%
m_b=4.4~GeV$, $\sin^2{\theta_{W}}=0.23117$, and $\alpha_{EW} = 1/128$\cite
{s12}.
\par
We take the MSSM parameters in the frame of the mSUGRA scenario. Of the five
input parameters ($m_0$, $m_{1/2}$, $A_0$, $\tan{\beta}$ and sign of $\mu$)
in this theory, we take $m_{1/2}$=120 GeV, $A_0$=300 GeV, $\mu>0$ and set
$\tan{\beta}$ to typical values. $m_0$ is tuned to obtain desired
$m_{H^{\pm}} $ which fall in the experimentally constrained range.
\par
In the numerical evaluation for the THDM case, we adopted the
following MSSM Higgs mass relations considering the radiative
corrections to fix the neutral Higgs boson masses and the mixing
angle $\alpha$ \cite{Esp}.
\begin{eqnarray} \label{alphah}
\tan{2 \alpha}&=& \tan{2 \beta}
    \frac{m_A^2 + m_Z^2}{m_A^2- m_Z^2+\epsilon/\cos(2\beta)},
    ~~ where~~ m_A^2 = m_{H^{\pm}}^2 - m_W^2,\\
m^2_{H^0,h^0}&=&\frac{1}{2} \left[ m_{AZ}^2 \pm \sqrt{m_{AZ}^4- 4
m_Z^2 m_A^2 cos^2 2 \beta-4\epsilon(m_A^2\sin^2\beta+
m_Z^2\cos^2\beta) } \right]
\end{eqnarray}
with
\begin{eqnarray} \label{hcorr1}
 \epsilon=\frac{3G_F}{\sqrt{2} \pi^2}\frac{m_t^4}{\sin^2\beta} \log
        \left[\frac{m_{\tilde{t}_L} m_{\tilde{t}_R} }{m_t^2} \right],
\end{eqnarray}
where the left-handed and right-handed scale top-quark masses are
determined by formulas in a grand unified (GUT) framework
\cite{Nath}
\begin{eqnarray} \label{hcorr2}
m_{\tilde{t}_L}^2&=& m_0^2+ \left(\frac{1}{2}
-\frac{2}{3}s_W^2\right)
                   \cos2\beta m_Z^2+ m_t^2, \nonumber \\
m_{\tilde{t}_R}^2&=& m_0^2+ \frac{2}{3} s_W^2 \cos2\beta m_Z^2+
m_t^2, \label{mstlr}
\end{eqnarray}
where at the Plank mass scale we assumed $ M_{\tilde{Q}}^2 =
M_{\tilde{U}}^2 =M_{\tilde{D}}^2=m_{0}^2$. That is to say in the
THDM case we take the charged Higgs mass $m_{H^{\pm}}$, $m_0$ and
$\tan\beta$ as input parameters and put their quantities being the
same as the corresponding ones in the MSSM case when they are
allowed in the mSUGRA parameter space, so as to make comparison of
the results from both models. We noticed also that there are some
charged Higgs mass regions in our plots which are not covered by
the allowed parameter space in the mSUGRA scenario(e.g. the dashed
lines for the THDM in Fig.3 where $\tan\beta=2$, $100~GeV \leq
m_{H^{\pm}} \leq 300~GeV$, and $m_{H^{\pm}}=200~GeV$, $2\leq
\tan\beta \leq 4$ in Fig.4a, et cetera), there we take
$m_0=220~GeV$. Actually our evaluation shows that by adopting
mSUGRA parameters and Eqs.(\ref{alphah}$\sim$\ref{mstlr}) for the
MSSM and THDM models respectively, we can get the corresponding
neutral Higgs masses and mixing angle $\alpha$ with approximately
same quantities in both models. To illustrate this point, we list
parameter quantities of neutral Higgs masses and mixing angle
$\alpha$ with $\tan\beta=6$ used in our calculations for both THDM
and MSSM models in table 1.

\begin{center} {\bf Table 1} The comparison of the corresponding
parameter values in the THDM and MSSM with $\tan\beta=6$\\
(the MSSM values from the mSUGRA/the THDM values from
Eqs.(\ref{alphah}$\sim$\ref{mstlr}))\\
\begin{tabular}{|c|c|c|c|c|}
\hline
$m_{H^{\pm}}[GeV]$ &$m_{h^0}[GeV]$&$m_{H^0}[GeV]$&$m_{A^0}[GeV]$&$\alpha$ \\
\hline
210    & {}98.50/96.54   & 197.40/197.12 & 194.60/193.99 & -0.2547/-0.2644 \\
300    & 100.10/97.45  & 291.09/290.83 & 289.30/289.02 & -0.2007/-0.2039 \\
450    & 102.09/97.80  & 444.06/443.86 & 442.75/442.76 & -0.1799/-0.1807  \\
550    & 103.25/97.88  & 545.15/544.98 & 543.95/544.09 & -0.1750/-0.1753  \\
\hline
\end{tabular}
\end{center}
\par
Fig.3 displays the integrated total cross section of
$W^{\pm}H^{\mp}$ production at LC with $\sqrt{s}=1~TeV$ versus the
mass of $H^{\pm}$ with $\tan{\beta}=2,6,32$, respectively. There
we can see the sophisticated structures on the MSSM curves, which
are induced mainly by the threshold effects from Yukawa couplings
of charged Higgs boson to quarks and scalar quarks, at the
vicinities where $m_{H^+}=m_{\tilde{t}_{i}}+m_{\tilde{b}_{i}}$
(i=1,2) and $m_{H^+}= m_{t}+m_{b}$ from loop diagrams\cite{glwh2}.
It is obvious that the supersymmetric particles in loop can
enhance the cross sections of the subprocess substantially by one
order. When the ratio of the vacuum expectation values has a small
value and the charged Higgs boson mass is in the mediate range,
the cross sections of the subprocess can reach $10^{-1}$ femto
barn. Since in the case of the MSSM we use the mSUGRA senario,
while in the THDM we use another parameter setting methode
mentioned above, some parts of parameter space with small
$m_{H^+}$ and $\tan\beta$ in the THDM are not allowed in the
mSUGRA parameter sapce. That is why we see in Fig.3 that the
curves for the MSSM are not plotted at the lower end.
\par
We present the cross sections of $W^{\pm}H^{\mp}$ productions at LC versus
$\tan{\beta}$ in Fig.4a and Fig4.b for a non-SUSY THDM and the MSSM,
respectively. Fig.4a is for $\sqrt{s}=1~TeV$, the $\tan{\beta}$ is in the
range of $2-35$ and the mass of the charged Higgs boson is taken as $200~GeV$,
$400~GeV$ and $600~GeV$, respectively. The Fig.4a shows that the cross
sections are reduced with the increment of the charged Hoggs boson mass in
both models. In Fig.4a, we can see that when the charged Higgs
boson has the mass value about $200~GeV$ and $\sqrt{s}=1~TeV$,
the cross sections at LC are in the order of $10^{-2}$ femto barn in the MSSM
THDM and in the range of $10^{-3} - 10^{-2}$ femto barn in the non-SUSY THDM.
Fig.4b is for $\sqrt{s}=500~GeV$ and the mass of the charged Higgs boson is
chosen as $200~GeV$. The two curves correspond to the results for the non-SUSY
THDM and the MSSM. All these curves in Fig.4a and Fig.4b show that the cross
section is a obviously effected by $\tan{\beta}$. That is because the couplings
of Higgs bosons to quark(squark) pairs are related to the ratio of the vacuum
expectation values. We can see from these figures that the dependence of the
cross section on $\tan{\beta}$ in the MSSM is weaker than in the non-SUSY THDM
and the maximal enhancement occurs at the low $\tan{\beta}$ range. We
compared our results in the non-SUSY THDM with those in Ref.\cite{eewh1} and
find that the production rates via $\gamma\gamma$ collision mode are of the
same order as those via $e^+ e^-$ collision mode at LC.

\section{Conclusion}
\par
In this paper, we study the loop induced $W^{\pm}$-associated production of
charged Higgs bosons via photon-photon fusion at linear colliders in the
non-SUSY THDM as well as in the MSSM. Numerical analysis of their production
rates is carried out at one-loop order with some typical parameter sets.
With our input parameters, the contributions from SUSY sector can lead to an
order of magnitude enhancement of the cross section for the case of the
THDM in some parameter space. Our results demonstrate that this production
rate via photon-photon collisions at LC can reach about 0.1 femto barn in
frame of the MSSM at LC when $m_{H^{\pm}} \sim 200~GeV$. We find that the
$H^{\pm}W^{\mp}$-associated production rate via $\gamma\gamma$ collisions
at LC is quantitatively comparable to that via the electron-positron collision
mode\cite{eewh1}.

\vskip 4mm \noindent{\large {\bf Acknowledgement:}}
\par
This work was supported in part by the National Natural Science
Foundation of China(project numbers: 19875049, 10005009), a grant from
the Education Ministry of China and the State Commission of Science and
Technology of China. The authors would like to thank
Dr. Zhang Ren-You for helpful discussions.

\renewcommand{\theequation}{A.\arabic{equation}}
\renewcommand{\thesection}{}
\section{Appendix}
\par
The Feynman rules we used for the couplings can be found in \cite{haber}. We
denote the couplings concerned in this work as:
\begin{eqnarray*}
V_{B^{+}tb}&=&V^{L}_{Btb}P_L + V^{R}_{Btb}P_R, \\
V_{B^{+}\tilde{t}_i \tilde{b}_j}&=&V^{L}_{B\tilde{t}_i \tilde{b}_j}P_L%
 + V^{R}_{B\tilde{t}_i \tilde{b}_j}P_R, \\
V_{B^{+}\tilde{\chi}^{0}_{i}\tilde{\chi}^{+}_{j}}&=& V^{L}_{B\tilde{\chi}%
^{0}_{i}\tilde{\chi}^{+}_{j}}P_L+ V^{R}_{B\tilde{\chi}^{0}_{i}\tilde{\chi}%
^{+}_{j}}P_R ~~~~(B^{+}=H^{+},G^{+}).
\end{eqnarray*}
\par
In our calculation there are several hundred loop diagrams contributing
to the subprocess $\gamma\gamma \to W^{\pm}H^{\mp}$. We find their
explicit expressions of one-loop diagrams are too complicated and
over lenthy to list here. Therefore, we list only the unrenormalized
self-energies expressions of mixing $H^+-W^-$ and $H^+-G^-$ explicitly
in appendix. Their fermion and boson parts are expressed as below:
\begin{eqnarray*}
\Sigma_{HW}^{f}(p^2)&=& -\frac{3i g d}{32 \sqrt{2} {\pi}^2} \left[ {m_t}
V_{Htb}^L B_0[p, m_t, m_b]+ ({m_t} V_{Htb}^L+{m_b} V_{Htb}^R) B_1[p, m_t,
m_b] \right] \\
&+& \sum_{i=1,2}^{j=1-4}\frac{1}{8{\pi}^2} \left\{ B_0[p, m_{\chi_{j}^0},
m_{\chi_{i}^+}] m_{\chi_{j}^0} ( V_{H{{\tilde{\chi}}_{j}^0}{{\tilde{\chi}}%
_{i}^+}}^{L*} V_{W{\tilde{\chi}}_{j}^0{\tilde{\chi}}_{i}^+}^L+ V_{H{\tilde{%
\chi}}_{j}^0{\tilde{\chi}}_{i}^+}^{R*} V_{W{\tilde{\chi}}_{j}^0{\tilde{\chi}}%
_{i}^+}^R) \right. \\
&+& \left. B_1[p, m_{\chi_{j}^0}, m_{\chi_{i}^+}] \left[ V_{H{{\tilde{\chi}}%
_{j}^0}{{\tilde{\chi}}_{i}^+}}^{L*} ( m_{\chi_{j}^0} V_{W{\tilde{\chi}}_{j}^0%
{\tilde{\chi}}_{i}^+}^L+ m_{\chi_{i}^+} V_{W{\tilde{\chi}}_{j}^0{\tilde{\chi}%
}_{i}^+}^R) \right. \right. \\
&+& \left. \left. V_{H{\tilde{\chi}}_{j}^0{\tilde{\chi}}_{i}^+}^{R*}
(m_{\chi_{i}^+} V_{W{\tilde{\chi}}_{j}^0{\tilde{\chi}}_{i}^+}^L+m_{%
\chi_{j}^0} V_{W{\tilde{\chi}}_{j}^0{\tilde{\chi}}_{i}^+}^R) \right] \right\}
\\
\Sigma_{HW}^{b}(p^2)&=& \frac{i g V_{H^+G^-h^0}}{32 {\pi}^2} (B_0[p,
m_{h^0}, m_W]+ 2 B_1[p, m_{h^0}, m_W]) \sin{(\alpha-\beta)} \\
&-& \frac{i g V_{H^+G^-H^0}}{32 {\pi}^2} (B_0[p, m_{H^0}, m_W]+ 2 B_1[p,
m_{H^0}, m_W] ) \cos{(\alpha-\beta)} \\
&-& \frac{i g V_{H^+H^-h^0}}{32 {\pi}^2} (B_0[p, m_{h^0}, m_{H^+}]+ 2 B_1[p,
m_{h^0}, m_{H^+}]) \cos{(\alpha-\beta)} \\
&+& \frac{i g V_{WWh^0}}{32 {\pi}^2} (B_0[p, m_{h^0}, m_W]- B_1[p, m_{h^0},
m_W]) \cos{(\alpha-\beta)} \\
&-& \frac{i g V_{H^+H^-H^0}}{32 {\pi}^2} (B_0[p, m_{H^0}, m_{H^+}]+ 2 B_1[p,
m_{H^0}, m_{H^+}]) \sin{(\alpha-\beta)} \\
&-& \frac{{m_W} g^2}{64 {\pi}^2} ( B_0[p, m_{H^0}, m_W]-B_1[p, m_{H^0},
m_W]) \sin{2(\alpha -\beta)} \\
&-& \sum_{i,j=1-2} {\left[ \frac{3 }{16{\pi}^2} B_0[p, m_{\tilde{t}_j}, m_{%
\tilde{b}_i}] V_{H\tilde{t}_j\tilde{b}_i} V_{W\tilde{b}_i\tilde{t}_j} +\frac{%
3}{8{\pi}^2 } B_1[p, m_{\tilde{t}_j}, m_{\tilde{b}_i}] V_{H\tilde{t}_j\tilde{%
b}_i} V_{W\tilde{b}_i\tilde{t}_j} \right] } \\
\Sigma_{HG}^{f}(p^2)&=& 3 d \left\{ -A_0[m_b] (V_{Htb}^L
V_{Gtb}^{L*}+V_{Htb}^R V_{Gtb}^{R*}) +p^2 B_1[p, m_t, m_b] (V_{Htb}^L \cdot
\right. \\
&& V_{Gtb}^{L*}+V_{Htb}^R V_{Gtb}^{R*})+{m_t} B_0[p, m_t, m_b] ({m_t}
V_{Htb}^L V_{Gtb}^{L*}+{m_b} \cdot \\
&& \left. V_{Htb}^R V_{Gtb}^{L*}+{m_b} V_{Htb}^L V_{Gtb}^{R*}+ {m_t}
V_{Htb}^R V_{Gtb}^{R*}) \right\}/(32 {\pi}^2) \\
&-& \sum_{i=1,2}^{j=1-4} \left\{ A_0[m_{\chi_{i}^+}] (V_{H{{\tilde{\chi}}%
_{j}^0} {{\tilde{\chi}}_{i}^+}}^{L*} V_{G{\tilde{\chi}}_{j}^0{\tilde{\chi}}%
_{i}^+}^L+ V_{H{\tilde{\chi}}_{j}^0{\tilde{\chi}}_{i}^+}^{R*} V_{G{\tilde{%
\chi}}_{j}^0{\tilde{\chi}}_{i}^+}^R) \right. \\
&& \left. -p^2 B_1[p, m_{\chi_{j}^0}, m_{\chi_{i}^+}] ( V_{H{{\tilde{\chi}}%
_{j}^0}{{\tilde{\chi}}_{i}^+}}^{L*} V_{G{\tilde{\chi}}_{j}^0{\tilde{\chi}}%
_{i}^+}^L + V_{H{\tilde{\chi}}_{j}^0{\tilde{\chi}}_{i}^+}^{R*} V_{G{\tilde{%
\chi}}_{j}^0{\tilde{\chi}}_{i}^+}^R) \right. \\
&& \left. -B_0[p, m_{\chi_{j}^0}, m_{\chi_{i}^+}] m_{\chi_{j}^0} \left[ V_{H{%
{\tilde{\chi}}_{j}^0}{{\tilde{\chi}}_{i}^+}}^{L*} (m_{\chi_{j}^0} V_{G{%
\tilde{\chi}}_{j}^0{\tilde{\chi}}_{i}^+}^L+m_{\chi_{i}^+} V_{G{\tilde{\chi}}%
_{j}^0{\tilde{\chi}}_{i}^+}^R) \right. \right. \\
&&+ \left. \left. V_{H{\tilde{\chi}}_{j}^0{\tilde{\chi}}_{i}^+}^{R*} (
m_{\chi_{i}^+} V_{G{\tilde{\chi}}_{j}^0{\tilde{\chi}}_{i}^+}^L+
m_{\chi_{j}^0} V_{G{\tilde{\chi}}_{j}^0{\tilde{\chi}}_{i}^+}^R) \right]
\right\}/(8 {\pi}^2) \\
\Sigma_{HG}^{b}(p^2)&=& (V_{H^+G^-h^0} V_{H^+H^-h^0} B_0[p, m_{h^0},
m_{H^+}])/(16 {\pi}^2) \\
&+& (V_{G^+G^-h^0} V_{H^+G^-h^0} B_0[p, m_{h^0}, m_W])/(16 {\pi}^2) \\
&+& (V_{H^+G^-H^0} V_{H^+H^-H^0} B_0[p, m_{H^0}, m_{H^+}])/(16 {\pi}^2) \\
&+& (V_{G^+G^-H^0} V_{H^+G^-H^0} B_0[p, m_{H^0}, m_W])/(16 {\pi}^2) \\
&-& \left\{ g^2 \left[ m_{H^+}^2 (B_0[p, m_{h^0}, m_W]-2 B_1[p, m_{h^0},
m_W] +B_{21}[p, m_{h^0}, m_W]) \right. \right. \\
&& \left. \left.-d B_{22}[p, m_{h^0}, m_W] \right] \sin{2(\alpha - \beta)}
\right\}/(128 {\pi}^2) \\
&+& \left\{ g^2 \left[ m_{H^+}^2 (B_0[p, m_{H^0}, m_W]- 2 B_1[p, m_{H^0},
m_W]+B_{21}[p, m_{H^0}, m_W]) \right. \right. \\
&& \left. \left. -d B_{22}[p, m_{H^0}, m_W] \right] \sin{2(\alpha -\beta)}
\right\}/(128 {\pi}^2) \\
&-& 3 \sum_{i,j=1-2}{} B_0[p, m_{\tilde{t}_j}, m_{\tilde{b}_i}] V_{G\tilde{b}%
_j\tilde{t}_i}^* V_{H\tilde{t}_j\tilde{b}_i}/(16{\pi}^2).
\end{eqnarray*}
\par
In case of the non-SUSY THDM, we take away the contributions from
supersymmetric particles, i.e. scalar quarks and charginos/neutralinos,
and used dimensional
regularization ($d=4-\epsilon$) instead of dimensional reduction (d=4).
As a check of our results, we find from our calculation that the amplitude
of all the diagrams in Fig.1 is UV-finite by themselves. The above
self-energy expressions for the non-SUSY THDM do not contradict with
those in Ref.\cite{eewh1}.
\par
In the above expressions we adopt the definitions of one-loop integral
functions in reference \cite{s13}. The numerical
calculation of the vector and tensor loop integral functions can be traced
back to four scalar loop integrals $A_{0}$, $B_{0}$, $C_{0}$, $D_{0}$ as
shown in \cite{passvelt}.

\vskip 10mm \begin{flushleft} {\bf Figure Captions} \end{flushleft}

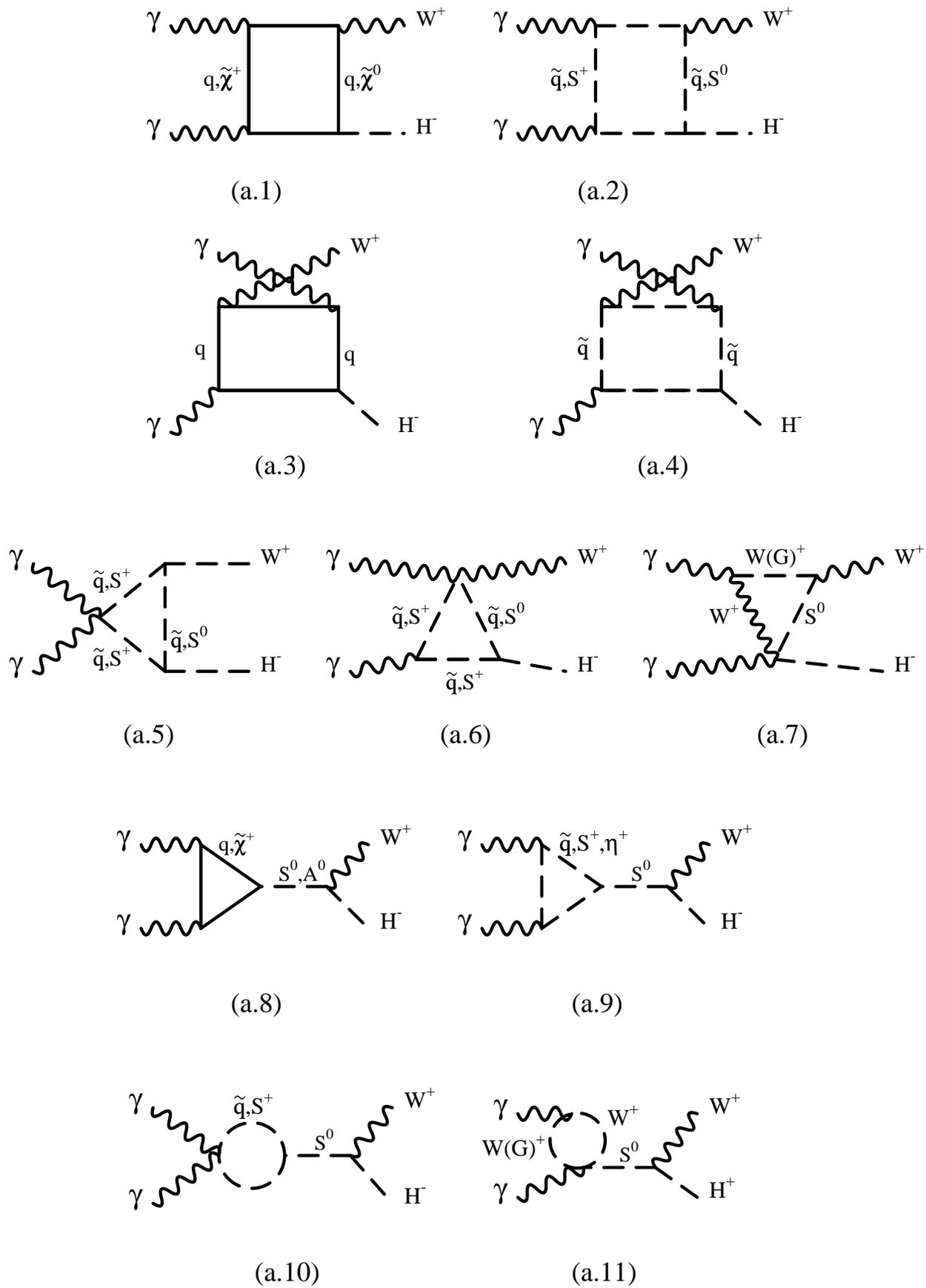
{\bf Fig.1} The Feynman diagrams of the subprocess $\gamma \gamma
\rightarrow W^{+}H^{-}$, where the notations $S^0$ and $S^+$ denote neutral
particles $h^0$($H^0$) and charged particles $H^+(W^+,G^+)$, respectively.

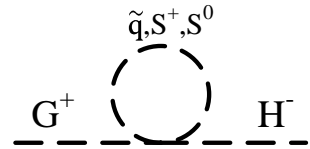
{\bf Fig.2} The diagrams contributing to self-energies of mixing $H^+-W^-$
(Fig.2(1-2)) and $H^+-G^-$ (Fig.2(3-5)).

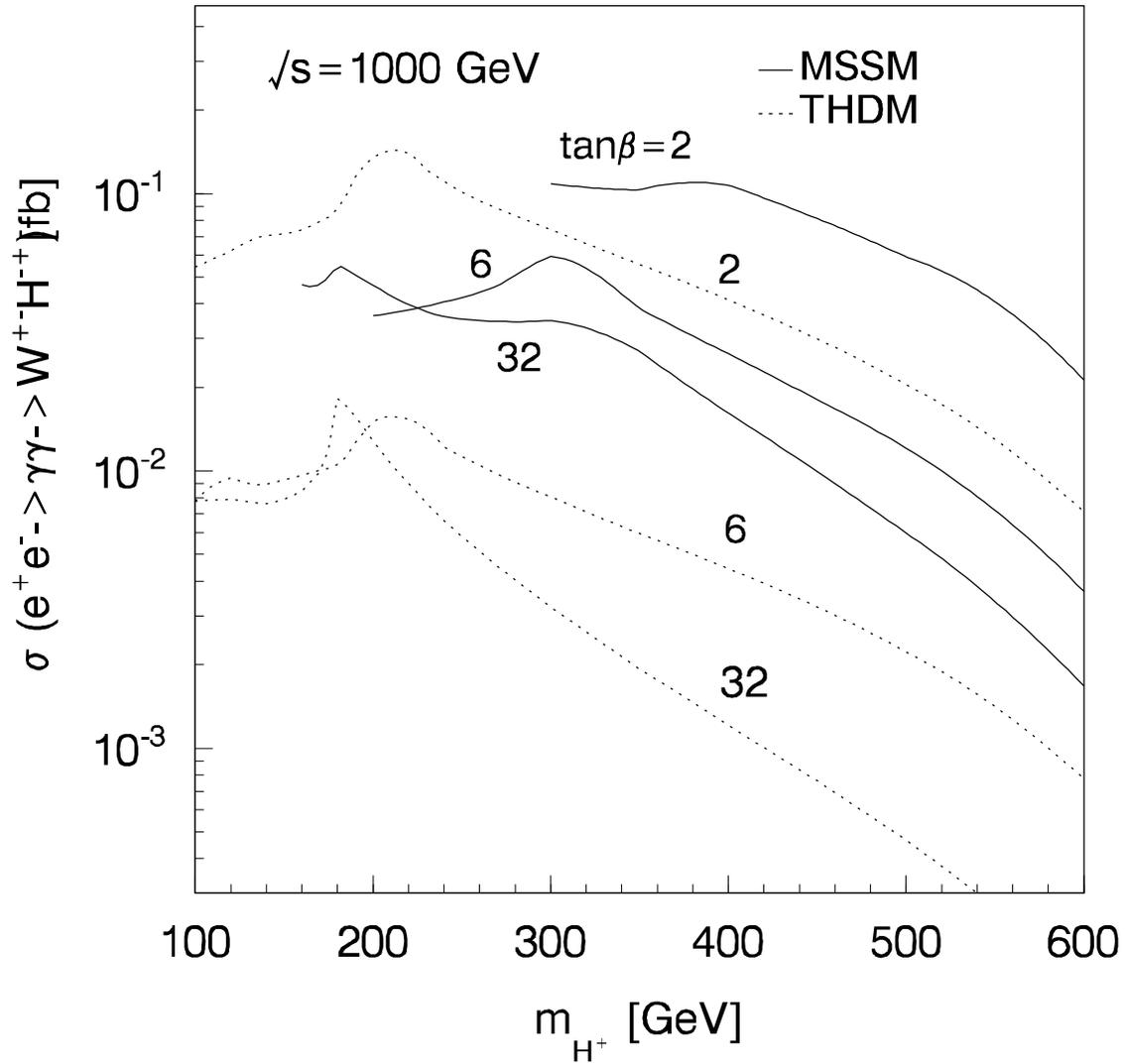
{\bf Fig.3} Total cross sections of $e^+e^- \to \gamma\gamma \to
W^{\pm}H^{\mp}$ in mSUGRA scenario as function of $m_{H^{\pm}}$ with $\sqrt{s%
}=1~TeV$ in non-SUSY THDM and the MSSM.

{\bf Fig.4a} Total cross sections of $e^+e^- \to \gamma\gamma \to
W^{\pm}H^{\mp}$ in mSUGRA scenario as function of $\tan\beta$ with $\sqrt{s}%
=1000~GeV$ in non-SUSY THDM and the MSSM.

{\bf Fig.4b} Total cross sections of $e^+e^- \to \gamma\gamma \to
W^{\pm}H^{\mp}$ in mSUGRA scenario as function of $\tan\beta$ with $\sqrt{s}%
=500~GeV$ in non-SUSY THDM and the MSSM.

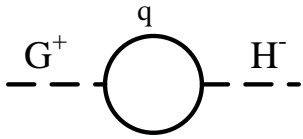
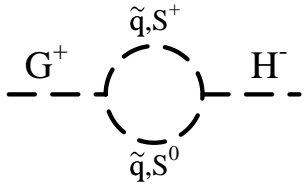
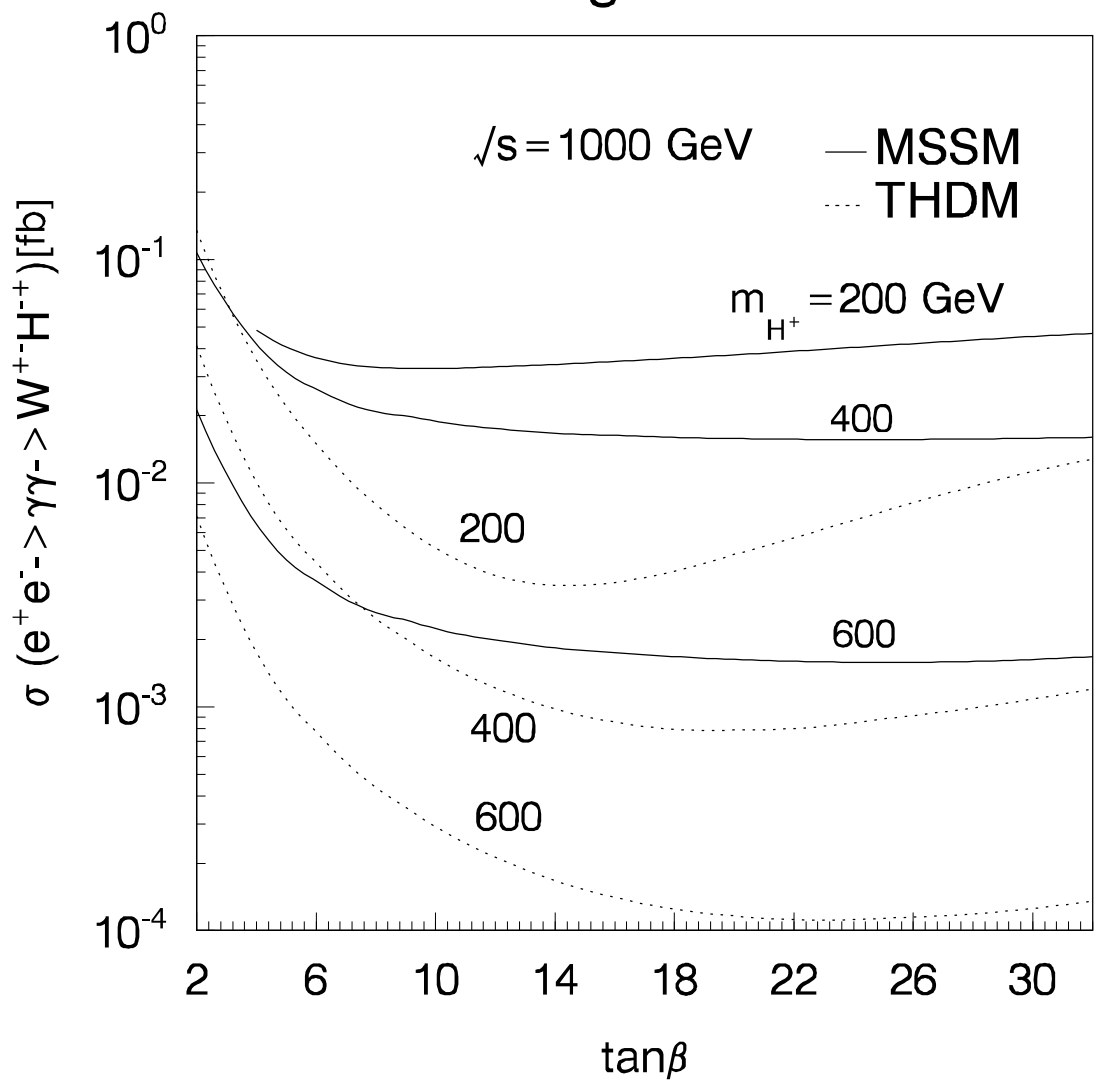
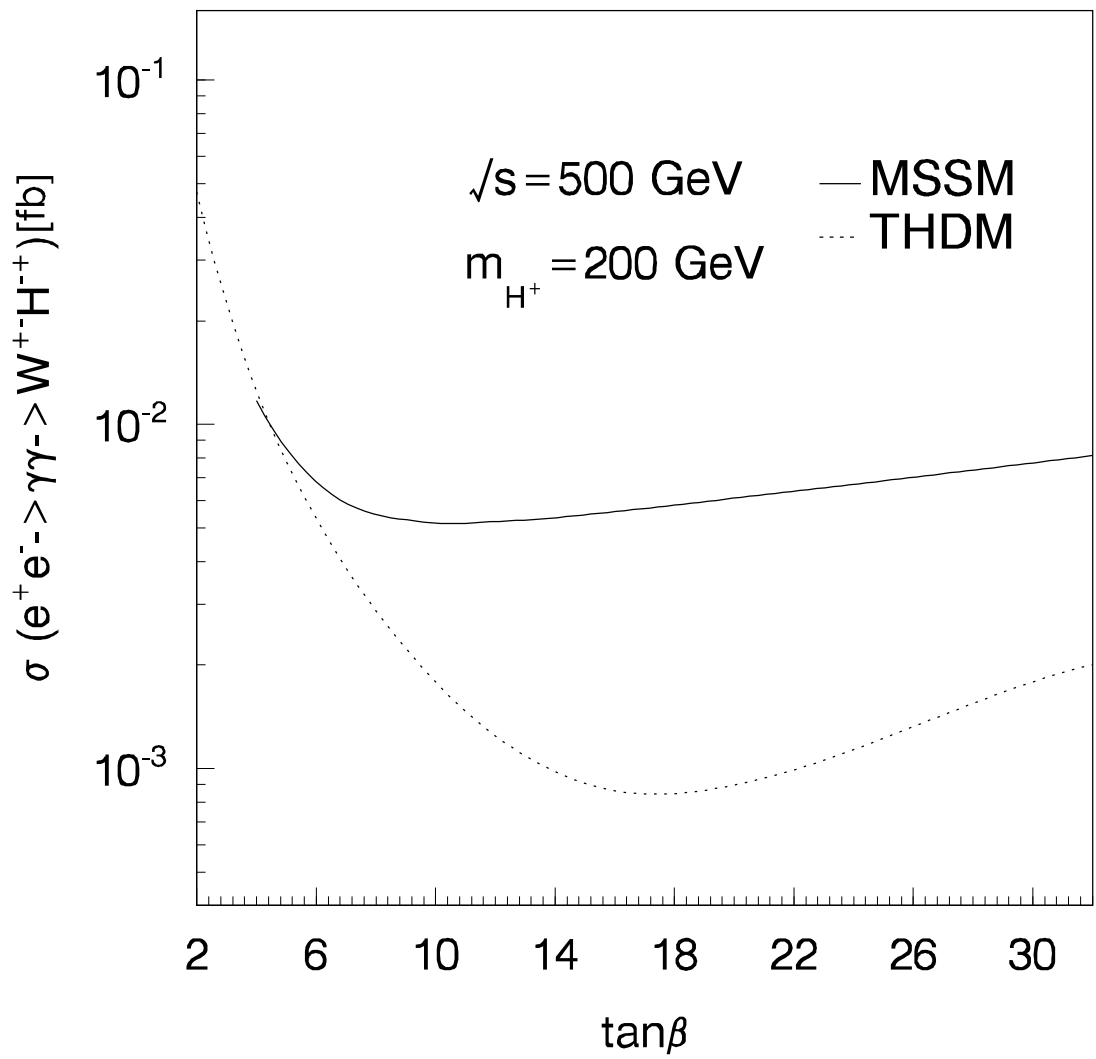
\end{document}